# Resilience and inequality growth monitoring after disaster using indicators based on energy production

Julien Gargani[1,2]

[1]Université Paris-Saclay, CNRS, Geops, Orsay, France

[2]Université Paris-Saclay, Centre d'Alembert, Orsay, France

**Highlights**

-Resilience and inequality growth between territories are analyzed using energy production

-Energy production changes after disasters affect socioeconomic activities' resilience

-Natural disasters favor inequality growth between territories during resilience

**Abstract**

The estimation of resilience and the determination of inequality growth after a disaster is often difficult. In this study, specific indicators were developed to analyze resilience, and the trajectories of resilience were compared. Energy production is relevant for describing variations in social and economic activities. This indicator was applied to a case study of a natural disaster, that is Hurricane Irma in 2017. The hurricane caused fatalities and destruction in the Caribbean islands of Saint Martin and Saint Barthelemy, which are two French overseas territories. Energy production after Hurricane Irma exhibited a significant decrease due to the destruction of the electricity network as well as perturbations in economic and social activities. The energy production restoration rate was higher in Saint Barthelemy than in Saint Martin. The electricity restoration rate after a disaster was almost constant. However, the energy production 18 months after Hurricane Irma was identical to that before Hurricane Irma in Saint Barthelemy; this was not the case in Saint Martin. During resilience, an increase in the gap between energy production in Saint Barthelemy and Saint Martin was observed. This study indicated that this gap represents an inequality growth between Saint Barthelemy (gross domestic product (GDP) of 40 000 euros/inhabitant) and Saint Martin (GDP of 16 600 euros/inhabitant). The indicators emphasized that inequality growth after natural disasters favors less vulnerable and more resilient territories. The number of inhabitants must be considered during indicator construction to avoid any bias.

## 1. Introduction

Disasters caused by extreme events may increase with climate change due to rising sea-level, and consequently, the intensity of extreme events increases [1]. Coastal areas are particularly affected by natural risk, and Caribbean islands are vulnerable to hurricanes [2,3,4]. In the case of a disaster, (i) infrastructure and buildings are physically damaged, (ii) social and economic activities are disturbed, and (iii) fatalities and injuries occur [4,5]. Risk management focuses on reducing the potential negative effects of disasters by anticipating these effects and proposing solutions. However, the quantification and monitoring of some of these data may be difficult after a disaster.

During the last decade, vulnerability and resilience have been used to describe and improve risk management. Vulnerability refers to the inability to withstand the effects of a hostile environment. Various types of vulnerabilities exist, such as social or environmental vulnerability, and they could be very specific, such as vulnerability to hurricanes or marine submersion. Vulnerability to natural disasters reveals the multidimensionality of natural disasters and their numerous and complex relationships with social and environmental scenarios [6]. Vulnerability is multidimensional, and the totality of relationships in a social or environmental scenario could be affected *a priori* by a change [6-Bankoff et al., 2004]. Numerous studies have focused on vulnerability and proposed quantification through specific indicators [7]. Nevertheless, quantification of vulnerability from a natural disaster is difficult because of the multidimensional aspects that must be considered and are not well described by a financial cost [3,5]. The vulnerability indicator should be a composite of multiple quantitative standardized indicators to incorporate the multidimensional aspects. Thus, comparisons could be made and coordinated mitigation achieved. However, vulnerability and/or economic indicators cannot be interpreted without considering the historical context [8-Armatte, 2010], and the quantification of vulnerability remains challenging.

Many different definitions of resilience have been proposed in different fields ranging from physics to psychology. It has been defined as a process of adaptation in the face of adversity, crisis, or stress [9]. Nevertheless, several authors assume that resilience is a trait and not a process. De Terte and Ian [10] defined resilience as having the ability to cope with a crisis or to rapidly return to the pre-crisis stage. When this concept is used to describe post-disaster evolution, it can be gauged on an individual, community, or physical level, for example, when the effects on infrastructure [11]. Economic resilience can be strengthened by implementing policies aimed at mitigating the risks and effects of severe crises. However, the benefits of economic policies must be balanced against their costs in terms of reduced expected growth.

Resilience has been criticized because this concept is often ambiguous and involves responding instead of promoting prevention. It could also promote injustice by favorably describing the scenario leading to a disruptive change that could have a political, social, or economic origin [12]. For example, circumstances that cannot be changed are not easily defined objectively and that, consequently, should be accepted. Policies of resilience can put the responsibility of disaster response on individuals rather than publicly coordinated efforts, and promoting resilience draws attention away from governmental responsibility and self-responsibility [11]. To consider the collective dimension of resilience, resilience can be analyzed at the territorial rather than the individual scale.

Vulnerability and resilience indicators can be considered scientific concepts, as well as government tools. Economic resilience anticipation can be ambiguous because transformation of society and adaptation of individuals after a disaster or crisis could be oriented without democratic procedure [13]. Creative destruction by economic crisis has been considered a fundamental criterion for economic dynamics in the past [14]. Nevertheless, changes are not necessarily positive after a crisis or disaster. In this study, no hypothesis was formed on the role of disasters in economic evolution. Herein, we considered that resilience is a trajectory that returns to equilibrium after the severe perturbation of a territory. Perturbation could be caused by environmental, economic, social, or sanitary changes. The trajectory of resilience corresponds to a temporal evolution of the state of every element that returns or does not return to its initial state before any perturbation.

More specifically, in this study, the perturbation corresponding to Hurricane Irma that occurred in 2017 was classified as five on the Saffir-Simpson scale, which caused damage, trauma, and deaths that affected the economic, social, sanitary, and environmental scenarios in the Caribbean [15,4]. Natural disasters cause damage to infrastructure, including electricity plants and networks, which can have serious effects on health [16] and economic and social activities [17]. The reduction or interruption of electricity influences the availability of other networks (water, communication, sewerage, etc.) [18]. Furthermore, a decrease in economic and social activities cause a reduction in energy demand [19,20].

Various activities has been affected by Hurricane Irma [5]. Quantifying all the changes in the activities impacted by Hurricane Irma with an accurate time step is difficult. Many social and economic activities are directly or indirectly associated with energy consumption [18]. The influence of energy on society has increased since the 19$^{th}$ century and has deeply transformed and constrained the social as well as economic activities at present [21]. Human-nature interactions are conditioned by techniques and, in particular, energy production [22]. Consequently, analysis of resilience of social and economic activities is difficult without considering the evolution of energy production or consumption. Thus, this study considered energy production as a key indicator for analyzing resilience.

Monitoring energy consumption allows the analysis of the evolution of social and economic activities. Electricity consumption depends on (i) the hour of the day, (ii) the day of the week, and (iii) the season [23]. Energy consumption changes when people are at work or at home, during social events or when temperature changes [24,23]. Previous studies have emphasized the relationship between wealth and energy consumption [25]. Although the effect of the COVID-19 pandemic on electricity generation has been accurately analyzed recently [26,27,28], the evolution of energy production after other kinds of disasters has been less explored [17]. The effect of hurricanes on energy production and consumption is often combined with other social and economic effects and has been rarely studied. Resilience analysis of two Caribbean islands after Hurricane Irma was performed by studying electricity generation. The possibility that environmental disasters increase inequality between individuals has been suggested by several studies [30], but remains unverified and has never been proposed for territories relying on electricity generation after a natural disaster. In this specific case, the electricity plant and network were damaged. The area considered in this study is composed of two islands, Saint Martin and Saint Barthelemy, which are two French overseas territories. Comparison of two different islands allowed the distinction of general trends related to major hurricanes from other specific influences due to local effects. The insular context has specific constraints on the energy system, such as supplementary cost in production (fuel transport) and consumption (water desalination) [29]. Electricity consumption in the two studied territories was monitored during the last 12 yr, which was used to analyze the temporal variations and correlated those with natural or anthropic events.

The research questions addressed by this study are as follows:

-the manner in which electricity generation evolves after a natural disaster and

-if an efficient indicator can be established based on electricity generation to characterize the variation in inequality between territories.

Moreover, this study aimed to discuss the following aspects:

-the efficiency of electricity generation in characterizing resilience after natural disasters and

-if disasters promote increase in inequality between territories.

## 2. Context and method

### 2.1. Context

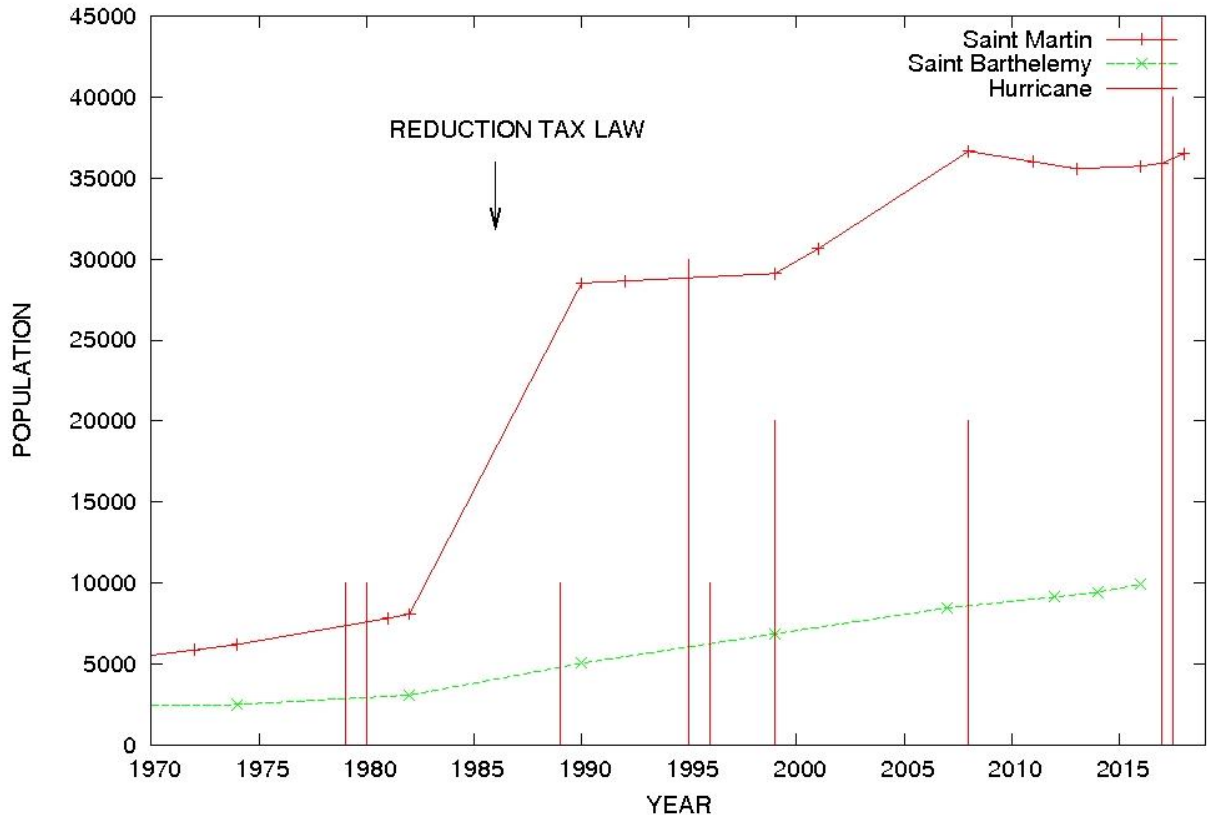

*FIGURE 1: Number of inhabitants in Saint Martin and Saint Barthelemy [3,17]. Hurricanes are indicated by vertical lines and their lengths are based on the Saffir-Simpson scale.*

Saint Martin and Saint Barthelemy are two French Caribbean islands with 35 000 and 10 000 inhabitants, respectively [31]. Their population has increased significantly during the last decade (Figure 1), after which the French government implemented the tax exemption law to develop tourism and economic activities [32,33]. In 2014, the gross domestic product (GDP) of Saint Martin and Saint Barthelemy were 16 600 euros/inhabitant and 40 000 euros/inhabitant, respectively [34,35].

The island of Saint Martin is divided between the French side, called Saint-Martin, and the Dutch side, called Sint-Maarten, both with approximately the same number of inhabitants. The area of Saint Martin size 53 km$^2$, while that of Saint Barthelemy is 24 km$^2$. The number of tourists arriving annually via air are 0.1-2 x 10$^6$ in Saint-Martin/Sint-Maartens and 0.3 x 10$^6$ in Saint Barthelemy [34,35]. Tourism represents the main economic activity of these islands [3,36], with tourists from North America and Europe.

The two studied territories have many similarities: (i) both are islands in the Caribbean, (ii) their economies are dominated by the tourism industry, (iii) both are French overseas territories, and (iv) their areas and number of inhabitants are small (<50 000 inhabitants; <100 km$^2$). However, they differ by their generated income, that is, their different per capita GDPs. Although GDP cannot accurately describe all social and economic activities [37], it suggests that the average income in Saint Barthelemy is higher than that in Saint Martin. Field work conducted in December 2019 and satellite image analysis revealed that the income is distributed more homogeneously in Saint Barthelemy than in Saint Martin, where low income is overrepresented in several neighborhoods (Sandy Ground, Orléans) [3].

Electricity in the two studied territories is mainly supplied by the Electricité de France (EDF). Electricity is generated by two power plants, one located in Saint Barthelemy and the other in Saint Martin. Fuel is used to produce electricity, and the production of renewable energy is negligible (2% in 2019); geothermal energy is not available locally, aeolian energy is not developed, and use of solar energy is slowly increasing. In Saint Barthelemy, the production capacity reaches 34.2 MW [38]. In Saint Martin, the power plant can produce 56.6 MW, whereas the peak of daily electricity generation is 27.5 MW [39]. Individual electricity generators are present in the better-off neighborhoods and used in the case of a blackout, but their total electricity generation is unknown.

## 2.2. Method

Yearly, monthly, and daily peak electricity production data in Saint Martin and Saint Barthelemy were used in this study. Annual electricity generation was available from 2006 to 2019 for Saint Martin [40,35] and from 2007 to 2019 [34]. For comparison, yearly electricity generation data from 2007 to 2019 was utilized, when the annual electricity generation was available for both the islands. Annual electricity generation ($E_Y$) was used to analyze long-term (i.e., decanal) evolution. The annual electricity generation was standardized using the value recorded in the 2007 ($E_{y2007}$). Specifically, standardized annual electricity generation was obtained by dividing electricity generation $E_Y$ by the value of annual electricity generation recorded in 2007. Standardized annual electricity generation (i.e., $E_{SY} = E_Y/E_{Y2007}$) evolution over time was compared for the two islands. The ratio between the standardized annual electricity generation of Saint Martin $E_{SY-SM}$ (SY-SM = standard yearly for Saint Barthelemy) and standardized annual electricity generation of Saint Barthelemy $E_{SY-SB}$ was used to analyze the relative evolution of electricity generation. This represented a case study to evaluate the efficiency of an indicator based on electricity generation to monitor the evolution of social and economic activities in the two different territories. In the general case, the indicator $I_Y = [E_{Y\text{-Territory 1}}(t)/E_{Y\text{-Standard Year-Territory 1}}]/[E_{Y\text{-Territory 2}}(t)/E_{Y\text{-Standard Year-Territory 2}}]$ measures the variation in electricity generation between the two territories.

There were two to eight years between the two different estimations. Demographic evolution was required to interpret trends in electricity consumption $E_Y$ over several years. To determine the effect of demographic variation on electricity generation, a specific indicator was constructed $I_{Y\text{-Demo}} = [E_{Y\text{-Territory 1}}/E_{Y\text{-Standard-Territory 1}}/\text{inhabitant}]/[E_{Y\text{-Territory 2}}/E_{Y\text{-Standard-Territory 2}}/\text{inhabitant}]$ and compared to $I_Y$. In the case of significant demographic variation, $I_{Y\text{-Demo}}$ could be more appropriate than $I_Y$ to measure the intensity of variation in electricity generation between the two different territories. The number of inhabitants in Saint Martin and Saint Barthelemy has been collected by the French National Institute of Statistics (INSEE) since 1954 [31]. These indicators were constructed to explore the discrepancies between Saint Barthelemy and Saint Martin. This indicator is based on the $E_{SY-SB}/E_{SY-SM}$ ratio. When this indicator is >1, it implies that social and economic activities that consume electricity have increased more in Saint Barthelemy (i.e., Territory 1) than in Saint Martin (i.e., Territory 2) and that the gap is further increasing between both islands.

Monthly electricity generation data in Saint Martin and Saint Barthelemy are available from January 2013 to May 2019 [34,35,38,39]. This study focused on the impact of Hurricane Irma that occurred in September 2017, and data on monthly electricity generation were presented and used from January 2017 to May 2019, over 29 months, including the cyclonic season. To compare monthly electricity generation between Saint Martin and Saint Barthelemy, a standardized value was obtained by dividing the monthly electricity generation ($E_M$) by the maximum value of monthly electricity generation ($E_{M-max}$). The maximum monthly electricity generation in this specific case is January 2017 for Saint Martin and March 2017 for Saint Barthelemy. The ratio between the standardized monthly electricity generation of Saint Barthelemy $E_{SM-SB}$ (SM-SB = standard monthly for Saint Barthelemy) and standardized monthly electricity generation of Saint Martin $E_{SM-SM}$ (SM-SM = standard monthly for Saint Martin) was used to compare electricity generation in Saint Martin and Saint Barthelemy. From a theoretical perspective, the indicator $I_M = [E_{M\text{-}Territory\ 1}/E_{M\text{-}Max\text{-}Territory\ 1}]/[E_{M\text{-}Territory\ 2}/E_{M\text{-}Max\text{-}Territory\ 2}]$ represented the variation in electricity generation between the two different territories.

To avoid any bias in the interpretation of monthly electricity generation due to different numbers of inhabitants in Territory 1 (in this study, Saint Barthelemy) and Territory 2 (in this study, Saint Martin), a comparison between monthly electricity generation per inhabitant for the two territories was also conducted. The comparison of monthly electricity generation per capita of Territory 1 ($E_{M\text{-}Territory\ 1}$/inhabitant) with that of Territory 2 ($E_{M\text{-}Territory\ 2}$/inhabitant) more accurately measured the intensity of the variation between the two different territories when a different number of inhabitants existed. It could also be useful when the consumption of electricity per capita differed significantly. The trend and absolute value of monthly electricity generation per inhabitant for Saint Martin $E_{M\text{-}SM}$/inhabitant and for Saint Barthelemy $E_{M\text{-}SB}$/inhabitant were analyzed. The use of the ratio of monthly electricity generation $E_M(t)$, such as $E_M(t)/E_{M\text{-}max}$ and $E_M(t)/E_{M\text{-}Max}$/inhabitant, instead of monthly electricity generation allowed us to more accurately compare the trends of the two different territories. Estimation of $E_M$/inhabitant was calculated considering that the number of inhabitants was constant for a period of 29 months at 35000 for Saint Martin and 10 000 for Saint Barthelemy. There are no official demographic estimates of the variation in the number of inhabitants between January 2017 and May 2019.

The long-term velocity of electricity restoration after disaster ($V_{LR}$) is defined by $V_{LR} = (E_{M\text{-}t}/E_{M\text{-}max} – E_{M\text{-}min}/E_{M\text{-}max})/\Delta t$, where $E_{M\text{-}max}$ is the maximum value of monthly electricity generation (in GWh), $E_{M\text{-}min}$ is the minimum value of monthly electricity generation after disaster (GWh), $E_{M\text{-}t}$ is the monthly electricity generation (GWh) at time $t$ (yr), and $\Delta t = t\text{-}t_0$ is the duration to restore monthly electricity generation, where $t_0$ is the time when a disaster occurs.

The peaks of daily electricity generation in Saint Martin and Saint Barthelemy are available before, during, and after Hurricane Irma [41]. These data present a better temporal accuracy than monthly and annual electricity generation, but for a shorter time. The peak of daily electricity generation represents the maximum electricity generation for a specific day. During April, the peak of electricity generation occurred at 19 h-20 h in Saint Martin and Saint Barthelemy when people used home appliances and lights.

$E_Y$ was obtained by summing the $E_M$ for 12 months: $E_Y = \sum_{n=1}^{n=12} E_M$. Moreover, monthly electricity generation was estimated by summing the daily electricity generation: $E_M = \sum_{n=1}^{n=30} E_D$. The relationship between daily electricity generation and peak of daily electricity generation is $E_{D\text{-}peak} = \max(E_D)$; however, it could be broadly approximated by $E_{D\text{-}peak} = \bar{E}_D/24 + 2\sigma$, where $\sigma$ is the standard deviation of daily electricity generation. $E_Y$, $E_M$, and $E_{D-peak}$ were directly collected from the literature [41,37,38].

## 3. Results

### 3.1. Observations

*3.1.1. Annual electricity production*

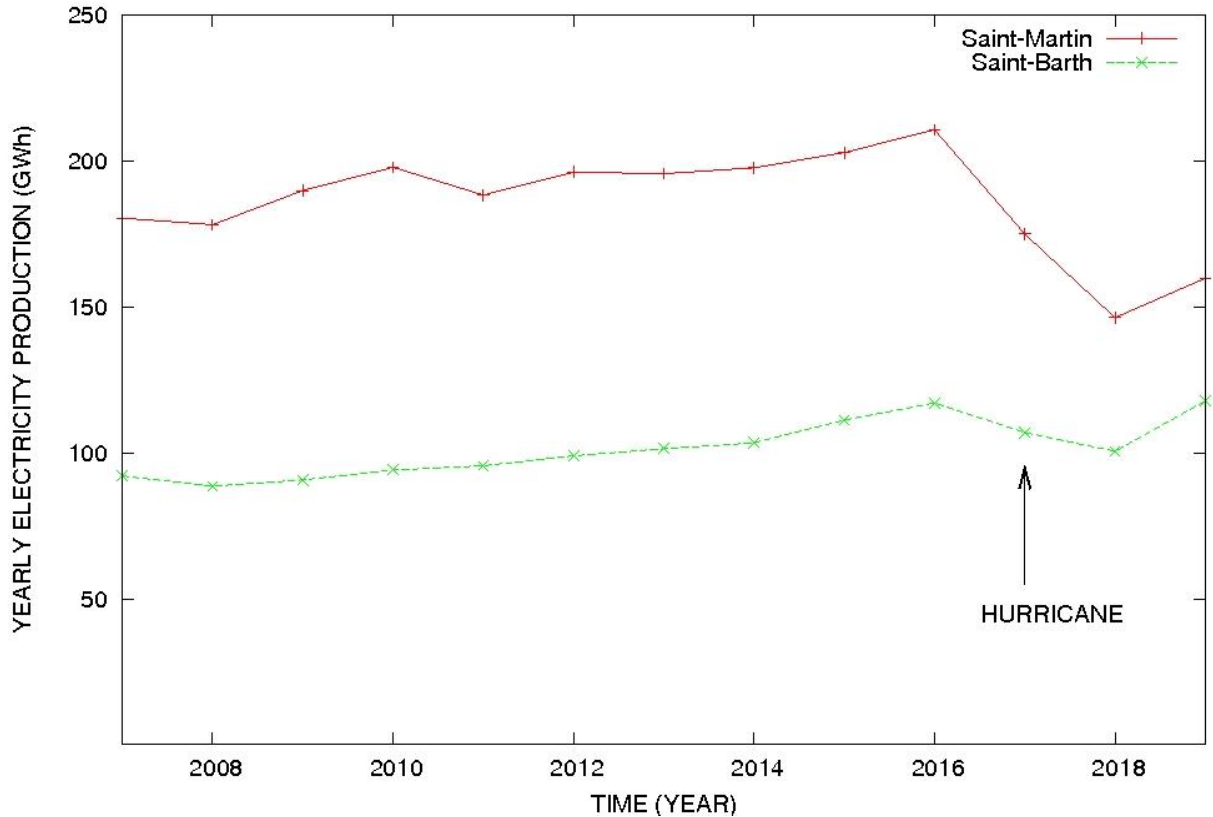

*FIGURE 2: Yearly electricity production in Saint Martin $E_{Y-SM}$ and Saint Barthelemy $E_{Y-SB}$ between 2007 and 2018. Data from [34,35,40].*

$E_{Y-SM}$ was between 30% and 50% higher than that of $E_{Y-SB}$ (Figure 2). The gap between the annual electricity generation of Saint Martin and Saint Barthelemy decreased and was minimal at approximately 30% in 2018.

From 2007 to 2016, a slow increase in the annual electricity generation occurred in Saint Barthelemy; $E_{Y-SB}$ increased from 90.6 GWh to 117.3 GWh. This represents an increase of +29.5% in nine years at a mean annual increase rate of approximately +3.3%. Then, a decrease of 14% in the annual electricity generation from 117.3 GWh in 2016 to 100.7 GWh in 2018 was observed. This represents a mean annual decrease rate of -8.3% in the annual electricity generation in Saint Barthelemy. In 2019, energy production increased in Saint Barthelemy and was higher than that in 2016 before Hurricane Irma.

In Saint Martin, $E_{Y-SM}$ increased from 180.6 GWh in 2007 to 210.8 GWh in 2016. This is a +16.7% increase in nine years, equivalent to an annual increase rate of +1.9%, which was lower than that in Saint Barthelemy. After 2016, $E_{Y-SM}$ decreased by 30.5% from 210.8 GWh to 146.5 GWh, representing an annual rate decrease of -15.25%, which was higher than that in Saint Barthelemy. In 2019, energy production increased in Saint Martin but remained lower than that in 2016.

### 3.1.2. Monthly electricity production

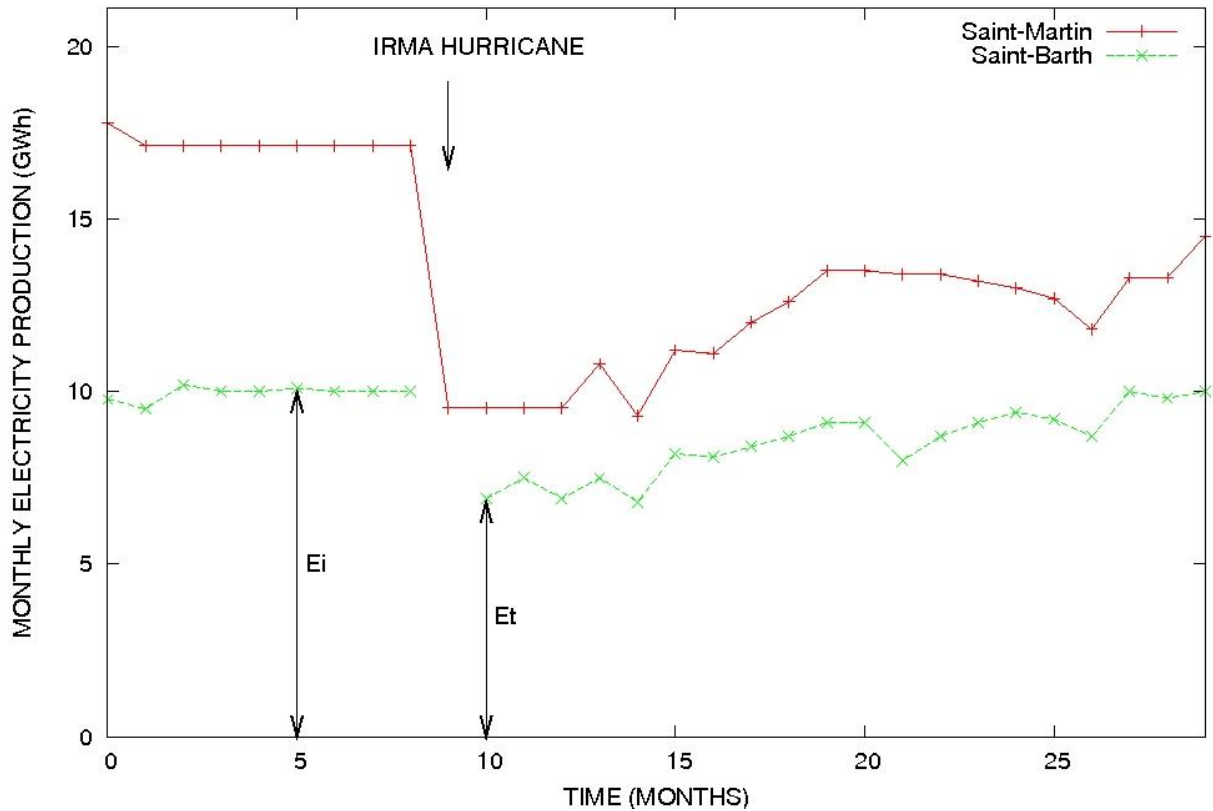

FIGURE 3: Monthly electricity production in Saint Martin $E_{M-SM}$ and Saint Barthelemy $E_{M-SB}$. Data from [34,35,40].

$E_{M-SM}$ was between 25% and 40% higher than $E_{M-SB}$ (Figure 3). The gap decreased from 40% to 25-30% corresponding to a decrease in energy production between Saint Martin and Saint Barthelemy of 10-15% after Hurricane Irma. In both cases, $E_M$ showed the same trend: an initial constant value, then an abrupt decrease to a low value that remained stable for 3-4 months, followed by a slow increase.

In Saint Barthelemy, the monthly electricity production $E_{M-SB}$ was 10 GWh during the first eight months of 2017. This initial value $E_{I-SB}$ =10 GWh was used as a reference for Saint-Barthelemy. Then, $E_{M-SB}$ decreased to 7 GWh in one month, representing a decrease of 30% in September 2017. During the next 19 months, a slow increase in $E_{M-SB}$ occurred from 7 to 10 GWh. Finally, a complete restoration of $E_{M-SB}$, identical to the values observed at the beginning of 2017, occurred in March 2019.

In Saint Martin, the monthly electricity production $E_{M-SM}$ was constant at approximately

17 GWh during the beginning of 2017 and decreased to 9.5 GWh in September 2017. This decrease represents 45% of the initial value of monthly electricity generation. The initial value $E_{I-SM}$ = 17 GWh was considered as a reference value for $E_{M-SM}$ in this study. The $E_{M-SM}$ increased from September 2017 (month 9) to May 2019 (month 29), where it reached a value of 14.5 GWh. In May 2019, $E_{M-SM}$ was always lower than the $E_{I-SM}$ value observed during the first eight months of 2017.

### 3.1.3. Peak daily electricity production

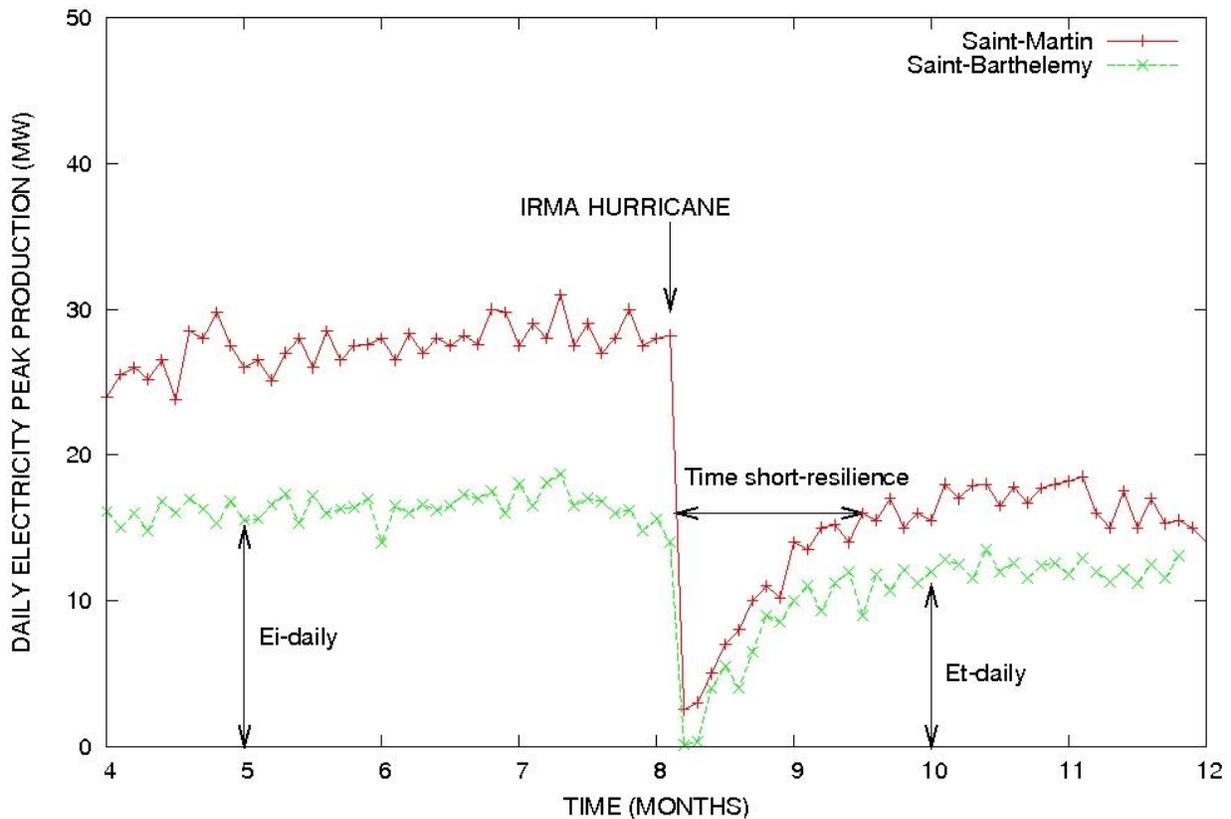

FIGURE 4: Daily electricity peak production in Saint Martin $E_{D-SM}$ and Saint Barthelemy $E_{D-SB}$. Data from [41].

The $E_{D-SM}$ in Saint Martin was between 30% and 40% higher than that of $E_{D-SB}$ (Figure 4). Moreover, the daily electricity peak production gap between Saint Martin and Saint Barthelemy was smaller after (approximately 30%) than before (approximately 40%) September 2017.

During several months before Hurricane Irma, $E_{D-SB}$ was almost constant at a value of 16.5 MW. Subsequently, there was an abrupt decrease in September 2017. During the first month following this decrease, an increase was observed up to 12 MW. After reaching 12 MW, the $E_{D-SB}$ became constant, and the daily electricity peak production was approximately 0.73% of the initial production value.

$E_{D-SM}$ was equal to 27.5 MW at the beginning, from May to August 2017. After an abrupt decrease of approximately 90% of $E_{D-SM}$ in September 5-6, 2017, an increase of daily electricity peak production up to 17 MW occurred 1.5-2 months later. Thereafter, the $E_{D-}$ $_{SM}$ was constant for several months at a value of 17 MW.

### 3.2. Interpretation

*3.2.1. Impact of hurricane on electricity generation*

To more accurately compare the trends of annual electricity generation of Saint Barthelemy ($E_{Y-SB}$) and that of Saint Martin ($E_{Y-SM}$), these values were divided by the annual electricity generation that occurred in 2007 in each case. In Saint Barthelemy, the standard annual electricity generation $E_{SY-SB} = E_{Y-SB} / E_{Y-SB-2007}$ increased steadily from 2008 to 2016, and decreased in 2017 and 2018. In Saint Martin, $E_{SY-SM}$ increased slowly compared to that in Saint Barthelemy and decreased more intensely in 2017 and 2018 (Figure 5).

The decrease in annual electricity generation in 2017 and 2018 was caused by Hurricane Irma, which occurred in September 2017. The hurricane destroyed the electricity network, and damaged buildings and infrastructure. In Saint Martin, Hurricane Irma damaged between 40% [4-Rey et al., 2019] and 90% [42-Gustin et al., 2018] of buildings. Tourism related infrastructure near the sea was particularly damaged by marine submersion [4-Rey et al., 2019]. The lower annual electricity generation in 2018 than that in 2017 could be because Hurricane Irma occurred in September 2017. During the first eight months of 2017, electricity generation was high, and the annual values represented a mean between the amount of electricity generated before and after the hurricane.

The peak of daily electricity generation confirmed that Hurricane Irma was responsible for the decrease in electricity generation (Figure 4). The power plant was stopped in Saint Martin after marine submersion affected the production unit. During 3-4 months, monthly electricity generation $E_M$ was low in Saint Martin and Saint Barthelemy. The apparent difficulty in restarting over several months was not solely due to delays in electricity network restoration, as the electricity network was almost completely restored after one month. However, the low $E_M$ was attributed to other factors, and it can be assumed that social and economic activities were not entirely restored.

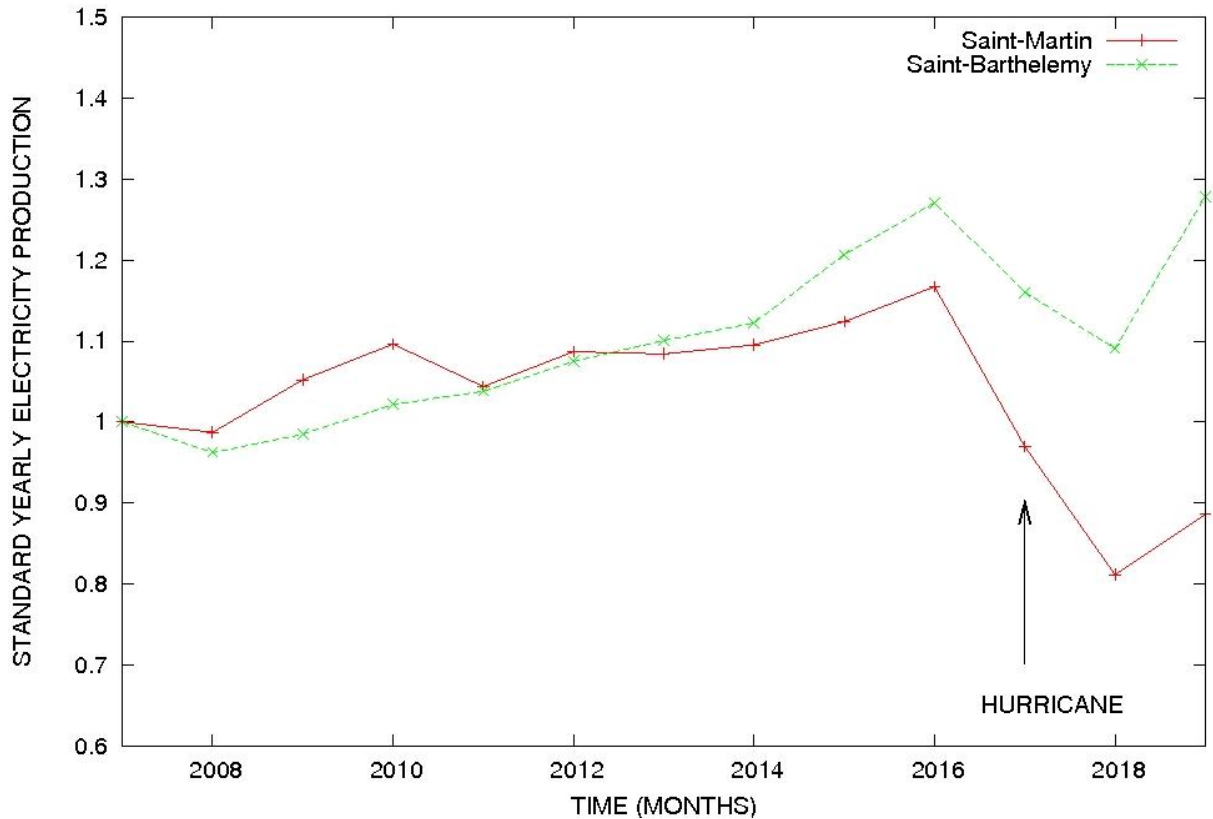

FIGURE 5: Evolution of annual electricity production of Saint Martin $E_{SY-SM}$ and Saint Barthelemy $E_{SY-SB}$. The annual electricity production was divided by electricity generated by each island in 2007.

*3.2.2. Electricity generation and resilience*

Standardized monthly electricity production $E_{SM}$ permits the comparison of Saint Barthelemy and Saint Martin in terms of electricity generation and resilience over 29 months (Figure 6). A slow increase in monthly electricity generation started after 3-5 months in both cases due to the gradual restoration of social and economic activities. In Saint Barthelemy, the monthly electricity generation was restored to the initial pre-disaster values after 18 months (Figures 3 and 6). This implied that restoration of economic and social activities at conditions equivalent to those that existed before Hurricane Irma were reached after $t_{long\ resilience}$ = 1.5 yr in Saint Barthelemy. Moreover, the long-term velocity of electricity restoration after disaster for Saint Barthelemy, $V_{LR-SB}$ = 0.2 yr$^{-1}$, considering that $E_t/E_{max}$ = 1, $E_{min}/E_{max}$ = 0.7, and $\Delta t = t_{long\ resilience}$ = 1.5 yr (table 1).

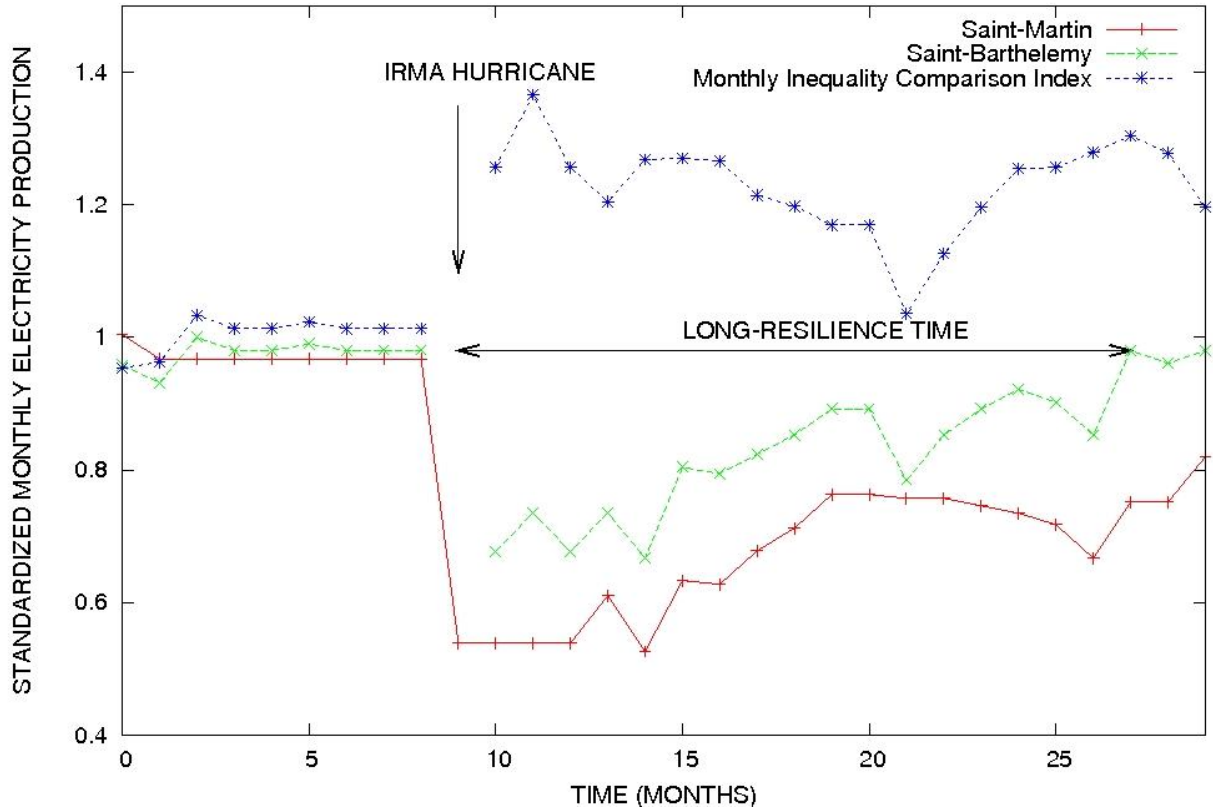

FIGURE 6: *Standardized monthly electricity production in Saint Martin $E_{SM-SM}$ and Saint Barthelemy $E_{SM-SB}$. Monthly electricity production from January 2017 to May 2019 was divided by the maximum value of monthly electricity production during this period in each island. A post-disaster relative inequality growth indicator was established from standard monthly electricity production $E_{SM-SB}/E_{SM-SM}$.*

In the case of Saint Martin, the restoration of standardized monthly electricity generation ($E_{SM-SM}$) was not been completed (Figure 6). The mean long-term electricity restoration velocity at Saint Martin $V_{LR-SM} = 0.15$ yr$^{-1}$ 20 months after Hurricane Irma, considering that $E_{SM-t}/E_{SM-max} = 0.8$, $E_{SM-min}/E_{SM-max} = 0.55$ (Table 1) and $\Delta t = 20$ months = 1.67 yr (Figure 6).

The long resilience time $t_{long\ resilience}$ required to restore the monthly electricity generation $E_{M-max}$ in Saint Martin as it was before Hurricane Irma was estimated by calculating $t_{long\ resilience} = (E_{M-max}/E_{M-max} - E_{M-min}/E_{M-max})/V_{LR-SM} = 3$ yr (September 2020). The long resilience velocity in Saint Martin was lower than that in Saint Barthelemy (Table 1). The electricity network was gradually restored in approximately one month, but electricity generation remained low. Thus, electricity network infrastructure was not responsible for the delays in electricity generation, and had other causes. The electricity network was improved after the hurricane, even though a part of the electricity network was buried.

| Island | GDP/inhab (€/inhabitant) | $E_i$ (GWh) | $t_{long\ resilience}$ (yr) | $E_{SM-MIN}/E_{SM-MAX}$ | $E_{SM-MAX} - E_{SM-MIN}$ (GWh) |
|---|---|---|---|---|---|
| Saint Martin | 16600 | 17 | 3 (estimated) | 0.55 | 0.45 |
| Saint Barthelemy | 40000 | 10 | 1,5 | 0.7 | 0.3 |

*Table 1: Main features of monthly electricity production.*

Resilience time, which is the time required to restore conditions to those before the disaster had occurred, was 1.5 yr in Saint Barthelemy and 3 yr in Saint Martin. A comparison of electricity generation in Saint Barthelemy and Saint Martin showed that when $E_{SM}$ was less impacted during 3-5 months just after a disaster, resilience occurred faster. In the present case, the smaller value of $E_{SM}$ in Saint Barthelemy was 15-20% higher than that in Saint Martin (Figure 6). This is due to the displacement of 7000-8000 inhabitants of Saint Martin from the island [42,43], which is discussed in the next section. The resilience velocity was also slightly higher in Saint Barthelemy than in Saint Martin due to the higher financial capacity of inhabitants in Saint Barthelemy, where reconstruction of damaged buildings was conducted using insurance funding as well as their financial stock. Building reconstruction favored the restart of economic activities. The activity of industrial harbors was characterized by an increase in concrete import during this period [34,35]. Moreover, tourism activities restarted more slowly.

*3.2.3. Impact of population change on electricity generation*

The growth in $E_{SY-SB}$ was higher from 2007 to 2016 than that in $E_{SY-SM}$ (Figure 5). This trend could be interpreted by considering a higher increase in social and economic activities in Saint Barthelemy over more than 10 yr. In this study, the change in the number of inhabitants was found to be responsible for a significant part of this trend. The number of inhabitants in a territory influences the number of social and economic activities. The increase in the number of inhabitants caused growth in the social and economic activities. During 2007-2016, population growth in Saint Barthelemy was higher (17%, from 8450 to 9912) than that in Saint Martin (<0.1%, from 35 714 to 35 746) and could explain a part of the higher growth of annual electricity generation observed in Saint Barthelemy.

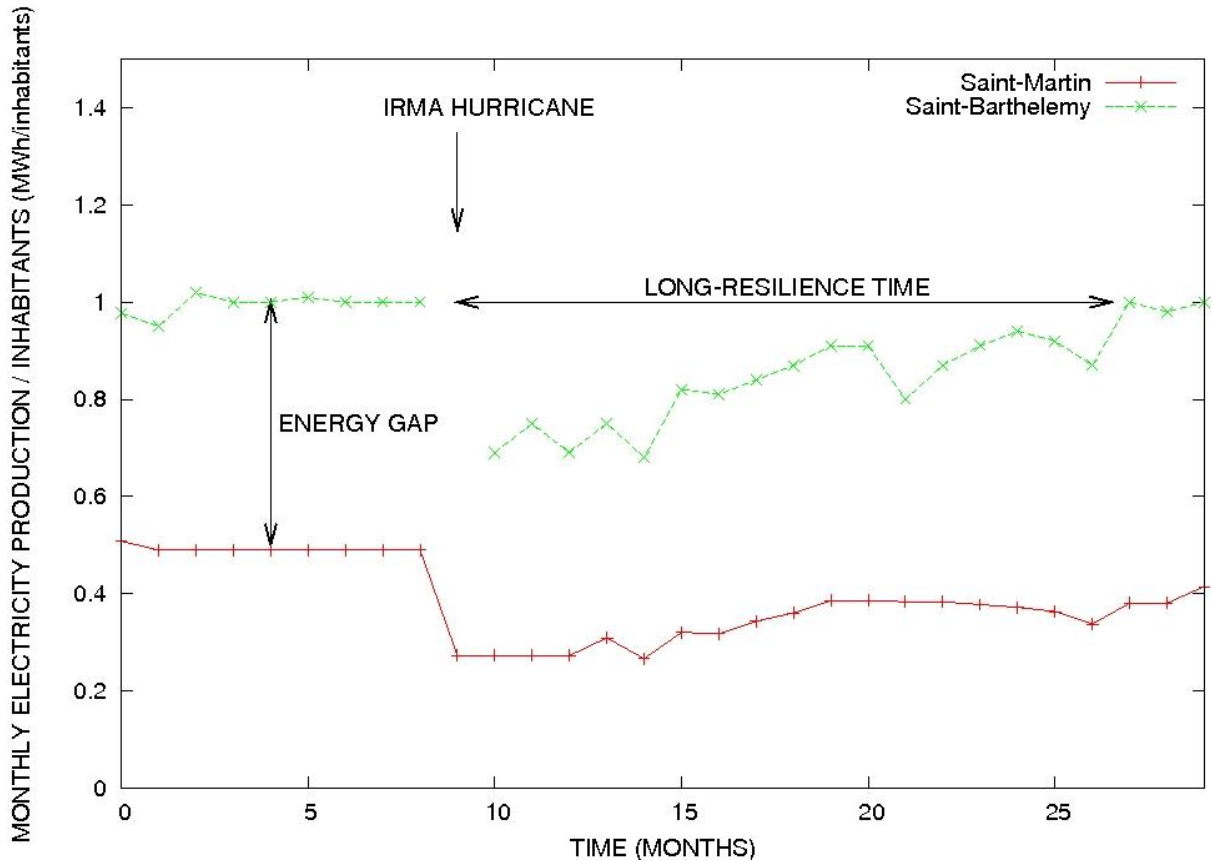

*FIGURE 7: Monthly electricity production/inhabitant in Saint Martin $E_{M-SM}$/inhabitant and Saint Barthelemy $E_{M-SB}$/inhabitant. The number is considered constant between 2017 and 2019 in Saint Martin (35000 inhabitants) and Saint Barthelemy (10000 inhabitants).*

The $E_M$ also depends on the number of inhabitants. It was higher in Saint Martin than in Saint Barthelemy (Figure 3). Figure 7 shows the monthly electricity generation per inhabitant, to avoid misinterpretation when comparing monthly electricity generation in Saint Martin and Saint Barthelemy. Monthly electricity generation per inhabitant was twice that in Saint Barthelemy than in Saint Martin (i.e., 1 MWh/inhabitant vs 0.5 MWh/inhabitant) before Hurricane Irma. The hurricane caused a decrease in the monthly electricity generation per inhabitant $E_M$/inhabitant, which was higher in Saint Barthelemy (0.3 MWh/inhabitant), than in Saint Martin (0.2 MWh/inhabitant). Nevertheless, these higher "absolute" impact must be interpreted carefully in term of vulnerability because those represent a decrease in the of monthly electricity generation/inhabitant of 30% (from 1 to 0.7 MWH/inhabitant) in Saint Barthelemy and that of 40% (from 0.5 to 0.3 MWh/inhabitant) in Saint Martin (Table 2). The lives of Saint Martin inhabitants were more affected than those of Saint Barthelemy by the disaster. Social and economic activities that consume electricity were more impacted in Saint Martin than in Saint Barthelemy. The energy gap $\Delta E$ between Saint Martin and Saint Barthelemy was 0.5 before Hurricane Irma (Figure 4) and increased up to 0.6. This represents an increase of 20% in the energy gap.

| Island | Number of inhabitants | $E_{M-i}$/inhab (GWh/inhab) | $E_{M-min}$/inhab (GWh/inhab) | $[E_{M-min}/inhab]/[E_{M-i}/inhab]$ | $[E_{M-i}/inhab]-[E_{M-min}/inhab]$ (GWh/inhab) |
|---|---|---|---|---|---|
| Saint Martin | 35000 | 0.5 | 0.3 | 0.6 | 0.2 |
| Saint Barthelemy | 10000 | 1 | 0.7 | 0.7 | 0.3 |

*Table 2: Main characteristics of electricity generation per inhabitant in Saint Martin and Saint Barthelemy.*

After Hurricane Irma, the population of Saint Martin decreased significantly (i.e., 7000-8000 inhabitants migrated from the island to Guadaloupe, France Metropolitan, or North America) during at least 4 to 6 months [42,43; Field interview]. This influenced the electricity consumption in Saint Martin and explained part of the discrepancy in the monthly electricity generation $E_M$ observed between Saint Martin and Saint Barthelemy just after the occurrence of the hurricane (3-5 months).

Population growth of approximately 10% in Saint Barthelemy caused an increase in electricity generation from 2007 to 2016 (Figure 5). In comparison, the number of inhabitants in Saint Martin was almost constant during this period and might have decreased. The influence of population growth differences between the two territories explained the discrepancy between the indicators $I_Y$ and $I_{Y-Demo}$. In particular, from 2017 to 2019 after Hurricane Irma, $I_Y$, and $I_{Y-Demo}$, where growth in $I_Y$ was approximately 25%, whereas $I_{Y-Demo}$ increased by approximately 20%. The indicator based on the number of inhabitants could be less accurate in 2017 because approximately 7000-8000 inhabitants left Saint Martin during several months after Hurricane Irma. Migration for several months could have reduced the recovery of electricity generation and resilience. Although the long-term population growth in Saint Barthelemy was not related to the effects of the disaster, the population evolution in Saint Martin in 2017 was due to the hurricane. The variation in the number of inhabitants after a disaster due to death or migration might occur and must be considered in the analysis of energy consumption.

*3.2.4. Relation between electricity generation and relative wealth*

The monthly electricity generation per inhabitant before Hurricane Irma $E_{M-i}$/inhabitant in Saint Barthelemy (1 GWh/inhab) was twice that in Saint Martin (0.5 GWh/inhab). There are no industries in the two islands, and the main economic activities are related to tourism (shop, restaurant, hotel, etc.). The number of arrivals at the Saint Barthelemy airport was estimated to be approximately 320 000 in 2017 [34]. The number of tourists arriving at the airports of Saint Martin/Sint-Maarten was 1.5-2 million [35] for approximately 73 000 inhabitants in the entire island in 2016-2017. Development of similar economic activities was observed for both the islands. Differences in electricity generation per inhabitant were not explained by the different economies but by different intensities of consumption.

Electricity generation in Saint Martin and Saint Barthelemy was due to consumption by electrical devices in luxurious residences (air conditioners, swimming-pool pumps, etc.), airport services, and building construction, but not by industry. In 2017, there were 0.74 buildings per capita in Saint Barthelemy, whereas there were only 0.34 buildings per capita in Saint Martin [44]. The GDP in Saint Barthelemy was more twice that in Saint Martin (Table 1). This implied that income in Saint Barthelemy was higher than that in Saint Martin. Consequently, a relationship could be considered to exist between income and electricity generation. This is in agreement with the findings of previous studies that showed that $CO_2$ production is higher in countries with higher GDP [45].

*3.2.5. Impact of hurricane on the increase in the gap in various activities*

Standard annual electricity generation was 8% lower in Saint Martin in 2016, 27% in 2018, and 40% in 2019 after Hurricane Irma. The standardized annual electricity generation $E_{SY}$ shows that the gap between Saint Barthelemy and Saint Martin was increasing, especially after Hurricane Irma during the resilience phase (Figure 5). The ratio between standard annual electricity generation in Saint Barthelemy and Saint Martin $E_{SY-SB}/E_{SY-SM}$ had been increasing since 2010 (Figure 8, red line) and had accelerated since 2017. This indicator revealed that since 2010, social and economic activities in which energy consumption is necessary have evolved and increased more in Saint Barthelemy than in Saint Martin. Thus, this indicator emphasized that differences (i.e., inequality) in social and economic activities that consume electricity was increasing between the two islands. Field observations conducted in December 2019 indicated that resilience was better in Saint Barthelemy than in Saint Martin, from a qualitative perspective. This is in agreement with the higher unemployment rate in Saint Martin than in Saint Barthelemy [34,35], as well as the higher reduction in tourists in Saint Martin than in Saint Barthelemy [34,35]. As indicator $I_Y$ represents an increasing difference in energy production between the two territories after a disaster, suggesting an increase in inequality between the two territories after Hurricane Irma, the indicator $I_Y$ can be considered reliable for detecting and characterizing the inequalities in other cases.

To avoid any bias in indicator construction, another indicator was established considering the population evolution $I_{Y-Demo}$ and it was compared with $I_Y$. This indicator (Figure 8, in green) incorporated the population evolution of each territory and exhibited the same trends as the previous indicator (in red). During the last decade, population growth had significantly increased in Saint Barthelemy and favored the development of social and economic activities. Consequently, electricity generation was also developed in Saint Barthelemy because of the population growth. The inequality growth in energy consumption between Saint Barthelemy and Saint Martin was favored by a higher population growth in Saint Barthelemy, and by the impact of Hurricane Irma on social and economic activities in 2017, 2018, and 2019 in Saint Martin. In 2017-2019, energy consumption was more impacted in Saint Martin than in Saint Barthelemy and caused an increasing gap.

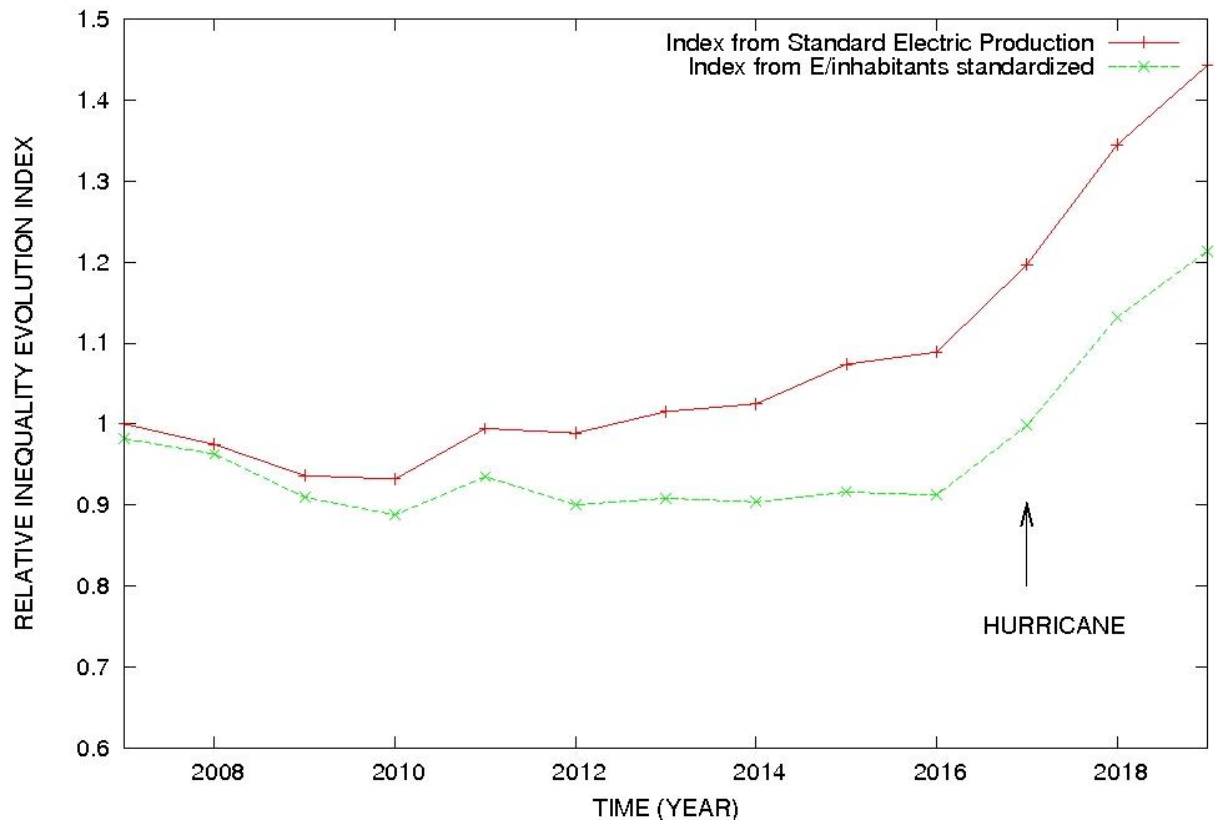

*FIGURE 8: Evolution of inequality with time. In red, the indicator is constructed with the ratio between the standard annual electricity production of Saint Barthelemy and Saint Martin, i.e. $E_{SY-SB}/E_{SY-SM}$. In green, the indicator in constructed using the ratio between the electricity per inhabitants standardized by electricity production in 2007 for Saint Barthelemy and for Saint Martin, i.e.* $[E_{Y-SB}/E_{Y-SB-2007}/inhabitant]/[E_{Y-SM}/E_{Y-SB-2007}/inhabitant]$.

Part of this increase in the gap was due to different population evolutions between Saint Barthelemy where it was increasing and Saint Martin, where it was constant during the last decade, except for a short decrease just after Hurricane Irma. The $E_{SY}$ also indicated that after Hurricane Irma, the gap continued to increase (Figure 6).

This also suggested that social and economic activities did not evolve at the same rate in Saint Barthelemy and Saint Martin. After Hurricane Irma, this ratio increased (Figure 6) and was always >1 until May 2019. Field analysis and interviews with stakeholders confirmed that reconstruction was more advanced in Saint Barthelemy than in Saint Martin in December 2019. Tourist flux also showed that the economy of Saint Barthelemy was at the same level as before Hurricane Irma [34].

The inequality growth between these two territories was due to (i) population growth of approximately 10%, as suggested by comparing the two curves in Figure 8 (one depending on population growth, the other independent of population growth), which was similar to the population growth of Saint Barthelemy corresponding to a stagnation in Saint Martin and (ii) to a different resilience after Hurricane Irma (the evolution of both $I_Y$ and $I_{Y-Demo}$ increased by more than 20% from 2016 to 2018).

## 4. Discussion

### 4.1. Influence of other events on electricity generation

The Irma disaster is not the only event whose impact can be analyzed with indicators based on electricity generation. During 2008-2010, a decrease in the indicators $I_Y$ and $I_{Y-Demo}$ was observed (Figure 8) in relation to the subprime mortgage crisis that started in the USA. North American tourists are numerous in the Caribbean, especially on these islands. Income related to North American (USA, Canada) tourists could be significant in the economic activities of Saint Martin and Saint Barthelemy [36,46,47,48]. Reduction of tourist flux [40,41] and electricity generation in these two islands occurred in 2008. Reduction of energy consumption in developed countries during financial crises has been described in previous studies [49]. The impact of this crisis was higher in Saint Barthelemy than in Saint Martin, as suggested by the decrease of 5-10% in $I_Y$ and $I_{Y-Demo}$ from 2007 to 2010. The financial income of inhabitants in these two islands might have been impacted differently due to different pre-existing income. Annual electricity generation increased more in Saint Martin than in Saint Barthelemy in 2009 and 2010, as shown in Figures 2 and 5. This increase was also observed in indicators $I_Y$ and $I_{Y-Demo}$, where a variation of 5% occurred. Contrary to the impact of Hurricane Irma, infrastructure was not directly impacted by the financial crisis. The difference between Saint Martin and Saint Barthelemy between 2007 and 2011 could also be due to a different economic strategy (mass vs. luxurious tourism), but could not be confirmed in this study.

The variation in electricity generation is not well understood. An increase in electricity generation of 10-20% occurred from 2014 to 2016 in both territories, corresponding to the period of higher number of tourists in the last 20 yr. The proportion of electricity consumption by tourists compared to that by residents is not well established in these territories. Their presence decreased during the monsoon from September to November and during the Covid-19 pandemic. The pandemic evidently influenced electricity consumption and generation, as observed in other areas [26,27,28], but the investigation of this effect is beyond the scope of this study.

During the recent decades, technological transformations such as numerical communication and platforms (Airbnb, etc.) and the increased utilization of air conditioners, electric cars, smartphones, and light-emitting diode (LED) home appliances have altered the electricity consumption trend. The difference in electricity consumption/inhabitant between the two territories was evident (Figure 7) and could be explained by the higher utilization of electrical devices in high-income territories. Nevertheless, from 2007 to 2018, a transformation that could increase the pre-existing differences at the same rate as the technological progress was difficult to establish.

### 4.2. Vulnerability and inequality

The direct or indirect effects of hurricane on health [16], stress [50], migration or displacement [52], and unemployment [34,35] have been described in previous studies. A heterogeneous distribution of the impact of disasters on inhabitants has been suggested in Saint Martin after Hurricane Irma [51] and in New Orleans after Hurricane Katrina [52]. This study demonstrated the occurrence of heterogeneous impacts at the scale of an island, not only at the individual scale. The decrease in standard annual electricity generation was higher in Saint Martin than in Saint Barthelemy in 2017 and 2018 after Hurricane Irma. During the first 4 months after Hurricane Irma in 2017, the difference of the electricity resilience ratio ($E_{t-daily}/E_{i-daily}$, see Figure 4) between Saint Martin and Saint Barthelemy can be explained by the migration of 7000-8000 inhabitants from Saint Martin [56]. The other differences can be explained by other processes. The primary reason for the increase in this gap was not directly determined through electricity generation analysis or the interpretation of the indicators $I_Y$ and $I_{Y-Demo}$. The results of this study can be considered as a starting point for the interpretation of processes that lead to inequality growth. Physical destruction caused by natural disasters has an impact on material goods. Income or wealth can be gained and accumulated via various modalities (buildings, money, stock options, cars, etc.), and material goods such as buildings or home appliances are only one modality to accumulate wealth. The destruction of wealth has not been homogeneously distributed because (i) a difference exists in the vulnerability of buildings between high-income and low-income groups [44], and (ii) high income groups could accumulate more wealth using financial stock.

Considering that the income of Saint Barthelemy was initially higher, this indicator emphasized the effect of a natural disaster on the inequality growth and reflected on the role of pre-existing inequalities on resilience. Moreover, a higher financial stock after a natural disaster could permit faster and better restart. The processes that play a role at the individual scale might produce an effect at the territorial scale. The precise relationship that causes socio-economic evolution in territories remains debatable [51,53,54].

Saint Martin was economically more vulnerable to hurricane than Saint Barthelemy, as suggested by the fact that the reduction of standardized monthly electricity generation caused by hurricanes was higher in Saint Martin ($E_{SM-min-SM} < E_{SM-min-SB}$, Figure 6) and less resilient ($V_{LR-SM} < V_{LR-SB}$). The higher vulnerability of Saint Martin might have been favored by the higher vulnerability of the inhabitants in that territory [44-]. Nevertheless, the difference in income is not the only factor that explains the variation in vulnerability between the two territories [55]. Prevention and rescue organizations can also reduce vulnerability. Organizational weakness in Saint Martin has been highlighted by previous studies and it plays an evident role in its vulnerability [15].

### 4.3. Comparison with other areas

Electricity production after hurricanes Irma and Maria in 2017 was also recorded in Puerto Rico [17]. The resilience time observed in monthly electricity production in Puerto Rico was 12 months [56-Gargani, in press] in comparison to 18 months in Saint Barthelemy. Considering that Puerto Rico has a lower GDP than Saint Barthelemy, it can be concluded that the resilience time was not proportional to the GDP in every case. This is also consistent with the fact that Cuba is evidently less vulnerable to hurricanes than other areas with higher GDP [55]. Nevertheless, financial support could favor (i) better infrastructure and building construction and (ii) increased funding for reconstruction [57]. Puerto Rico is a bigger island than Saint Martin and Saint Barthelemy; however, other factors play a role in the restoration of Puerto Rico as it has a more diversified economy [58,59,60]. When economic activities depend on a unique sector that is highly affected by hurricane destruction, economic activity slowly evolves and restoration requires longer [61-Rhiney, 2018]. Tourism related infrastructure (hotels, guest-houses, and restaurants) require longer to be restored than other infrastructure (airports and harbors), networks (electricity and telecom), industry, and agriculture when numerous failures occur, and market dynamics are low. Resilience is not only due to the delay in electrical network restoration. Restoration of social and economic activities also depends on other parameters, such as availability of restored buildings, funding, security, public relations with tourism agencies, and communication about the resilience of the territory [62,63,64,65,66].

### 4.4. Efficiency of the Indicators and their limitations

The ability to use electricity generation to characterize social, environmental, and economic events was suggested by the present case study. As previously described, the decrease in energy (electricity) production in Saint Martin and Saint Barthelemy indicated that social and economic activities decreased in these islands after Hurricane Irma. This is consistent with the simultaneous growth in unemployment in Saint Martin (+18,5%) and Saint Barthelemy [34,35], as well as with the decrease in arrivals at the airports (-25.9% in Saint Martin) [34,35], which characterized the decrease in economic activity in 2018. Tourist arrivals in Saint Barthelemy in December 2019 were similar to those in December 2016 [35-IEDOM, 2020], which suggested that economic activities had been restored. In contrast, tourist arrivals in Saint Martin in December 2019 remained lower than before hurricane.

Electricity generation was relevant to describing the resilience of social and economic activities. The slow increase in electricity generation after the collapse in September 2017 is consistent with an increase in tourist arrivals and an increase in import of concrete from September 2017 to 2019 [34,35]. Resilience monitoring with $E_M$/inhabitant permits the characterization of the time to return to the initial state more accurately than other indicators, such as seasonal tourist arrivals (Figure 7). The return to the initial state does not imply that each activity is identical to that before the disaster, but that the total energy consumption would be unchanged. When the $E_M$/inhabitant became identical to that before disaster, it implied that the activities consuming electricity continued to consume an equivalent quantity of electricity as before, but each activity had not been restored to its state before the disaster. Energy production is a *proxy* for various activities. The identity of indicators based on energy is a consequence of not only the identity of each activity, but also the consumption from the various activities.

During the post-disaster period, resilience and inequality growth were widely constant according to the $E_{M-Territory\ 1}/E_{M-Territory\ 2}$ ratio, thereby indicating that the scenario of Saint Barthelemy was consistently better than that of Saint Martin (Figure 6). Some variations could be due to seasonal activities related to tourism, which depended on the weather.

Comparison of the curves of indicators $I_Y$ and $I_{Y-Demo}$ (Figure 8) revealed the similar trends, and showed that the amplitude of variations might differ. Both indicators suggested that inequality growth was concentrated during the post-disaster period (2017-2019). The evolution of indicators was broadly similar, but after 12 yr, the evolution varied by more than 20%. A comparison of relative inequality indicators based on standardized energy showed that population variation could play a role in these indicators and must be neglected during interpretation. Any uncertainty in population variation could cause an uncertainty in the interpretation of the indicators. Considering the evolution of inhabitants in the indicators allowed us to confirm that part of the long-term economic growth in Saint Barthelemy was influenced by the increase in the number of local inhabitants and not only by the presence of a different organization. Information about the local context is required to accurately interpret the variation in indicators.

Inequality variation at the individual scale was beyond the scope of this study. Previous studies have suggested that inequalities increase after a disaster at the individual scale [13,30] or at the neighborhood scale [44,67,68,69,70], but not at the scale of an island. This suggests that inequality growth during the resilience phase after a natural disaster is observed at various scales (individuals, neighborhoods, and islands). The interdependence and causal relations between one scale and the others may be determined, but have not been investigated herein. This study was based on the energy (electricity) production recorded by the main electricity company in these islands. The individual generation of electricity was not considered. Individual generators of electricity are often used in these islands to avoid outage, and this production could be significant just after hurricane, when blackout occurs. Our analysis was conducted during 2007-2019 and the potential increase of renewable energies or strategies to mitigate energy consumption in the context of global warming were considered, but could play a role on electricity generation in the future [65].

**4.5. Disaster repetition and inequality growth**

The analysis of a recent disaster in this study revealed that a single event can cause an inequality growth of more than 20% between the two territories. Consequently, the repetition of natural disasters in the same area could lead to an accumulation of inequalities. This could represent a significant difference in electricity generation between the two territories after several disasters. For example, if an Irma-type-disaster occurred twice in the same territories, the inequality growth should be more than 40% and the electricity generation difference should be more than 20% when the number of disasters increases. At least 17 hurricanes have occurred during the last century in these territories [32]. We suggest that the 50% difference in electricity generation per capita between Saint Martin and Saint Barthelemy could be partly due to past disasters. Other economic processes or political decisions might have played a role in this difference, but the above-mentioned cause cannot be neglected. Disasters must be better considered in the analysis of inequality growth and the mitigation of natural hazards.

This case study showed that energy production or consumption is useful for monitoring and understanding the evolution of social and economic activities. Energy is relevant to analyzing the dynamics of societies because of its role in many activities. The effects of energy production on the environment influence political decisions in many countries. For example, strategies that favor sustainability modify future energy production to mitigate $CO_2$ production [45]. This study emphasized that the heterogeneous effects of natural disasters on territories impacted by the same event should be better anticipated in the future.

**5. Conclusion**

Investigating the energy production in terms of electricity generation in the two territories impacted by a natural disaster (Hurricane Irma) led to the following conclusions:

(1) Energy (electricity) production can be used to monitor social and economic activities.

(2) Disaster causes a significant decrease in electricity generation.

(3) Energy (electricity) production restoration rates differ across territories, even if these territories have been impacted by the same event.

(4) The energy restoration rate after a disaster is almost constant, and complete resilience could be anticipated.

(5) During resilience, an increase in the gap between electricity generation per inhabitant occurs between territories.

(6) Inequality growth occurs between territories during the resilience phase.

(7) The financial crisis that occurred in 2007-2008 caused a decrease in electricity generation in Saint Martin and Saint Barthelemy, and this effect was more intense in Saint Barthelemy than in Saint Martin.

(8) The high number of tourists during 2014-2016 caused an increase in energy (electricity) production.

(9) Population variation plays an evident role in electricity generation and must be considered in the analysis.

**Acknowledgement:** This work was supported by the French National Agency for Research ANR RELEV. The funding source was not involved in the design and interpretation of the data in the writing of the report. Kelly Pasquon helped with the field missions and interviews conducted in 2019.

**Declaration of interest:** none